\documentclass[twocolumn,pra,superscriptaddress]{revtex4}
\usepackage[dvipsnames,usenames]{color}
\usepackage{graphicx}
\usepackage{dcolumn}
\usepackage{bm}
\usepackage{amssymb}
\usepackage{amsmath}
\usepackage{extarrows}
\usepackage{color}

\begin{document}

\title{Exploring topological double-Weyl semimetals with cold atoms in optical lattices}

\author{Xue-Ying Mai}
\affiliation{Guangdong Provincial Key Laboratory of Quantum Engineering and Quantum Materials,
SPTE, South China Normal University, Guangzhou 510006, China}

\author{Dan-Wei Zhang}
\email{zdanwei@126.com}\affiliation{Guangdong Provincial Key Laboratory of Quantum Engineering and Quantum Materials,
SPTE, South China Normal University, Guangzhou 510006, China}

\author{Zhi Li}
\affiliation{Guangdong Provincial Key Laboratory of Quantum Engineering and Quantum Materials,
SPTE, South China Normal University, Guangzhou 510006, China}

\author{Shi-Liang Zhu}
\email{slzhu@nju.edu.cn} \affiliation{National Laboratory of Solid
State Microstructures and School of Physics, Nanjing University,
Nanjing 210093, China} \affiliation{Guangdong Provincial Key Laboratory of Quantum Engineering and Quantum Materials,
SPTE, South China Normal University, Guangzhou 510006, China} \affiliation{Synergetic Innovation Center of Quantum Information and Quantum
Physics, University of Science and Technology of China, Hefei
230026, China}

\begin{abstract}
We explore the topological properties of double-Weyl semimetals
with cold atoms in optical lattices. We first propose to realize
a tight-binding model of simulating the double-Weyl semimetal with
a pair of double-Weyl points by engineering the atomic hopping in
a three-dimensional optical lattice. We show that the double-Weyl
points with topological charges of $\pm2$ behave as sink and source
of Berry flux in momentum space connecting by two Fermi arcs and
they are stabilized by the $C_{4h}$ point-group symmetry. By
applying a realizable $C_4$ breaking term, we find that each
double-Weyl point splits into two single-Weyl points and obtain
rich phase diagrams in the parameter space spanned by the
strengths of an effective Zeeman term and the $C_4$ breaking term,
which contains a topological and a normal insulating phases and
two topological Weyl semimetal phases with eight and four
single-Weyl points, apart from the double-Weyl semimetal phase.
Furthermore, we demonstrate with numerical simulations that (i)
the mimicked double- and single-Weyl points
can be detected by measuring the atomic transfer fractions after a
Bloch oscillation; (ii) the Chern number of different quantum phases
in the phase diagram can be extracted from the center shift of the
hybrid Wannier functions, which can be directly measured with the time-of-flight
imaging; (iii) the band topology of the $C_4$-symmetric Bloch Hamiltonian can be detected simply from measuring the spin polarization
at the high symmetry momentum points with a condensate in the optical lattice.
The proposed system would provide a promising platform for elaborating
the intrinsic exotic physics of double-Weyl semimetals and the
related topological phase transitions.
\end{abstract}

\date{\today}

\maketitle

\section{introduction}

Recently, topological Weyl semimetals have attracted a broad
interest due to their wide range of exotic properties that are
distinct  from those of topological insulators
\cite{Kane,Qi,Balents,Burkov,Delplace,Zhao,Taas,WeylTheo1,SMExp1,SMExp2,Zyuzin,Parameswaran}.
Most importantly, the long-sought Weyl fermions, which are
massless spin-1/2 particles in quantum field theory but have never
been observed as fundamental particles in nature, can emerge as
gapless quasiparticle excitations near band touching points dubbed
as Weyl points in Weyl semimetals. The topological nature of the
Weyl points in three-dimensional momentum space supports the
existence of nontrivial Fermi arc surface states. The Weyl
fermions in the bulk and the Fermi arc states in the surfaces are
expected to give rise to exotic phenomena in Weyl semimetals, such
as anomalous electromagnetic responses
\cite{Zhao,Zyuzin,Parameswaran}. A significant advance has been
theoretically and experimentally made for exploring Weyl physics
not only in real materials \cite{WeylTheo1,SMExp1,SMExp2}, but
also in some artificial systems, such as photonic and acoustic
crystals \cite{Lu2013,Lu2015,Xiao2015,Chen2016}. Interestingly, a
new three-dimensional topological semimetal state, dubbed
double-Weyl semimetal, has been theoretically proposed in solids
that possess certain point-group symmetries
\cite{Multi,Srsi,Xu,Shivamoggi,Jian,Detect,Thermo,Lepori}. The
standard Weyl semimetal has linear dispersion in all the three
momentum directions near the single-Weyl points with topological
charge of $\pm1$. In contract, the dispersion of a double-Weyl
semimetal is quadratic in two dimensions and linear in the third
dimension near the gapless points with topological charge of
$\pm2$, which are thus named as double-Weyl points. Several materials have been proposed to be potential candidates
of double-Weyl semimetals \cite{Multi,Srsi,Xu,Shivamoggi,Jian,Detect,Thermo,Lepori}, and the
double-Weyl points have recently been observed in photonic crystals \cite{Chen2016}.
Some important properties of this newly predicted topological
state are rarely explored, such as the
symmetry-breaking effects and the topological phase transition.
Thus, other experimentally tunable systems for exploring the exotic topological properties double-Weyl semimetals are highly desirable.

On the other hand, ultracold atoms in optical lattices play an
important role in advancing our understanding of condensed matter
physics \cite{ColdAtom1}. Remarkably, as recent experimental
advances in realizing spin-orbit coupling and artificial gauge
field for neutral atoms \cite{GaugeRMP,GaugeRPP,SOC-Review}, these
systems provide a powerful platform with unparalleled
controllability towards studying topological states of matter. For
instance, the celebrated Harper-Hofstadter \cite{HHModel} model
and Haldane model \cite{HaldaneModel} have been realized
\cite{Miyake,Bloch2013a,Jotzu,Bloch2015,Shao} experimentally in
optical lattices, where the Chern number has also been
successfully probed. The chiral edge states have been
experimentally observed in one-dimensional optical lattices
subjected to a synthetic magnetic field and an artificial
dimension \cite{Cold-Edge1,Cold-Edge2}. The topological
(geometric) pumping has been demonstrated with cold atoms in
optical superlattices \cite{Pumping1,Pumping2,Pumping3}. The
two-dimensional spin-orbit coupling for Bose-Einstein condensates
has been realized in optical lattices and the band topology has
also been measured \cite{2DSOC}. Then an important question is
raised: can we realize other predicted topological phases that are
rare in solid-state materials in the cold atom systems? Several
proposals have been suggested to realize exotic topological
insulting states
\cite{Liu2013,Liu2014,Duan2014,Osterloh,Mazza,Goldman} and
topological nodal semimetals with single-Weyl points or nodal
loops \cite{Jiang,Dubcek,ZDW2015,He,Xu1,ZDW2016,Xu2,Shastri} in
optical lattices. Notably, it was proposed to
simulate the double-Weyl semimetals with ultracold atoms in
optical lattices in the presence of synthetic non-Abelian SU(2)
gauge potentials \cite{Lepori}. Other feasible schemes for
mimicking tunable double-Weyl semimetal states and detecting their
intrinsic topological properties in cold atomic systems are still
badly awaited.

In this paper, we explore the topological double-Weyl semimetals
with cold atoms in optical lattices. We first propose to realize
a tight-binding model of simulating the double-Weyl semimetal with
tunable double-Weyl points by engineering the atomic hopping in a
three-dimensional cubic optical lattice. We show that a pair of
double-Weyl points with nontrivial monopole charges behave as sink
and source of Berry fluxes in momentum space and they are stabilized
by the $C_{4h}$ point-group symmetry. We further investigate the
topological properties of the double-Weyl semimetal by calculating
$k_z$-dependent Chern number and the gapless edge states. By
applying a realizable $C_4$ breaking term, we find that each
double-Weyl point splits into two single-Weyl points and obtain a
rich phase diagram in the parameter space spanned by the strengths
of an effective Zeeman potential and the $C_4$ breaking term,
which contains a topological insulator phase, a normal band
insulator phase, and two topological Weyl semimetal phases with
eight and four single-Weyl points apart from the double-Weyl
semimetal phase. Finally, we demonstrate with numerical
simulations that (i) the analogous double- and single-Weyl points
can be detected by measuring the atomic transfer fractions after a Bloch
oscillation; (ii) the $k_z$-dependent Chern number of different
quantum phases in the phase diagram can be extracted from the
center shift of the hybrid Wannier functions, which are based on
time-of-flight imaging; (iii) the band topology of the $C_4$-symmetric Bloch
Hamiltonian can be detected simply from
measuring the spin polarization at the high symmetry momentum points
with a condensate in the optical lattice. The proposed cold-atom system
provides a promising platform for elaborating the intrinsic exotic
physics of double-Weyl semimetals and the related topological phase
transitions.

The paper is organized as follows. Section II introduces the
tight-binding model and optical-lattice system for realizing
double-Weyl semimetals with double-Weyl points. In Section III,
with the numerical calculation of the Chern number and the chiral
edge states, we elaborate the topological properties of the
simulated double-Weyl semimetals and obtain a rich phase diagram
containing other topological quantum phases by applying a symmetry
breaking term. In Section IV, we propose realistic schemes to
detect the simulated Weyl points and the characteristic
topological invariant with cold atoms in the optical lattice.
Finally, a short conclusion is given in Sec.V.

\section{model and system}
We consider a non-interacting (pseudo)spin-1/2 degenerate
fermionic gas in a three-dimensional cubic optical lattice, where
the spins are  encoded by two atomic internal states labeled as
$|\uparrow\rangle$ and $|\downarrow\rangle$. The tight-binding
Hamiltonian of the cold atom system is considered to be
\begin{eqnarray}
\hat{H}&=&\frac{t}{2}\sum_{\boldsymbol{r}}\left(\hat{a}_{\boldsymbol{r}+\boldsymbol{x},\uparrow}^{\dag}\hat{a}_{\boldsymbol{r},\downarrow}
-\hat{a}_{\boldsymbol{r}+\boldsymbol{y},\uparrow}^{\dag}\hat{a}_{\boldsymbol{r},\downarrow}+\text{H.c.}\right) \nonumber \\
&&-\frac{it}{4}\sum_{\boldsymbol{r}}\left[\hat{a}_{\boldsymbol{r}+(\boldsymbol{x}+\boldsymbol{y}),\uparrow}^{\dag}\hat{a}_{\boldsymbol{r},\downarrow}
-\hat{a}_{\boldsymbol{r}+(\boldsymbol{x}-\boldsymbol{y}),\uparrow}^{\dag}\hat{a}_{\boldsymbol{r},\downarrow}+\text{H.c.} \right]\nonumber \\
&&-\frac{t}{2}\sum_{\boldsymbol{r},\boldsymbol{\eta}}\left(\hat{a}_{\boldsymbol{r}+\boldsymbol{\eta},\uparrow}^{\dag}\hat{a}_{\boldsymbol{r},\uparrow}-\hat{a}_{\boldsymbol{r}-\boldsymbol{\eta},\downarrow}^{\dag}\hat{a}_{\boldsymbol{r},\downarrow}+\text{H.c.}\right)\\ &&+m_z\sum_{\boldsymbol{r}}\left(\hat{a}_{\boldsymbol{r},\uparrow}^{\dag}\hat{a}_{\boldsymbol{r},\uparrow}-\hat{a}_{\boldsymbol{r},\downarrow}^{\dag}\hat{a}_{\boldsymbol{r},\downarrow}\right), \nonumber
\end{eqnarray}
where $\hat{a}_{\boldsymbol{r},\sigma}$
($\hat{a}_{\boldsymbol{r},\sigma}^{\dag}$) is the annihilation
(creation) operator  on site $\boldsymbol{r}$ for the fermion with
spin $\sigma=\{\uparrow,\downarrow\}$, $\boldsymbol{\eta} =
\boldsymbol{x,y,z}$ denote the hopping directions, $m_z$ is the
strength of an effective Zeeman potential, and the hopping
strength is set $t=1$ as the energy unit hereafter. By defining
the two-component annihilation operator at site $\boldsymbol{r}$
as
$\hat{a}_{\boldsymbol{r}}=(\hat{a}_{\boldsymbol{r},\uparrow},\hat{a}_{\boldsymbol{r},\downarrow})^{T}$,
Hamiltonian (1) can be rewritten as
\begin{eqnarray}
\hat{H}&=&\sum_{\boldsymbol{r},\boldsymbol{\eta}}\left(\hat{a}_{\boldsymbol{r}+\boldsymbol{\eta}}^{\dag}U_{\eta}\hat{a}_{\boldsymbol{r}}+\text{H.c.}\right)+m_z\sum_{\boldsymbol{r}}\hat{a}_{\boldsymbol{r}}^{\dag}\sigma_z\hat{a}_{\boldsymbol{r}}\\
&&+\sum_{\boldsymbol{r}}\left[\hat{a}_{\boldsymbol{r}+(\boldsymbol{x}+\boldsymbol{y})}^{\dag}U_{xy}\hat{a}_{\boldsymbol{r}}-\hat{a}_{\boldsymbol{r}+(\boldsymbol{x}-\boldsymbol{y})}^{\dag}U_{xy}\hat{a}_{\boldsymbol{r}}+\text{H.c.}\right]\nonumber,
\end{eqnarray}
where the hopping matrices along the three axis are
$U_x=\frac{1}{2}(\sigma_x-\sigma_z)$,
$U_y=-\frac{1}{2}(\sigma_x+\sigma_z)$  and
$U_z=-\frac{1}{2}\sigma_z$, and along the $xy$ direction is
$U_{xy}=-\frac{1}{4}\sigma_y$, with $\sigma_{x,y,z}$ being the
Pauli matrices acting on the spin states.

Here the atomic hoppings $U_{\eta}$ and $U_{xy}$ between two
lattice sites along the  corresponding direction can be
spin-conserved hopping (the $\sigma_z$ term) or spin-flip hopping
(the $\sigma_x$ and $\sigma_y$ terms), which can be achieved by
the laser-assisted tunnelling technique with well-designed Raman
coupling between the two spin states
\cite{GaugeRMP,GaugeRPP,SOC-Review}. First, one can use a moderate
magnetic field to distinguish the spin states with the Zeeman
splitting, which allows one to correlate tunnelling in a spatial
direction with rotations in internal spin states and
state-dependent tunnelling phases. Second, the natural hopping
along each direction is suppressed by titling the cubic optical
lattice with a homogeneous energy gradient along the
$x,y,z$-directions, with the large tilt potential $\Delta
_{\eta}\gg t_{N}$ (such that the hopping probability
$\left(t_{N}/\Delta _{\eta}\right) ^{2}$ induced by the natural
tunneling is negligible) and $t_N$ denoting the natural tunneling
rate. The tilt potential can be created through the natural
gravitational field or the gradient of a dc- or ac-Stark shift,
and here we require different linear energy shifts per site
$\Delta_{x}\neq\Delta_{y}\neq\Delta_z\neq\Delta_{x}\pm\Delta_{y}$
in order to distinguish between the tunnellings directed along
different directions. Finally, the hopping terms can be restored
and engineered by application of two-photon Raman coupling with
the laser beams of proper configurations through the
laser-frequency and polarization selections
\cite{GaugeRMP,GaugeRPP,SOC-Review}. In principle,
arbitrary $2\times2$ hopping matrices including the required
$U_{\eta,xy}$ can be generated in this way with well-designed
laser configurations \cite{Osterloh,Mazza,Goldman}. Since several
protocols for implementing similar atomic hopping matrices and the
tunable Zeeman potential have been theoretically proposed or
experimentally realized
\cite{2DSOC,Liu2013,Liu2014,Duan2014,Osterloh,Mazza,Goldman} and a
different model of double-Weyl semimetals in optical lattices has
been presented, here we leave some details of realization out and
focus on exploration and detection of their novel topological
properties in the following.

\begin{figure}[tbph]
\centering
\includegraphics[width=8.5cm]{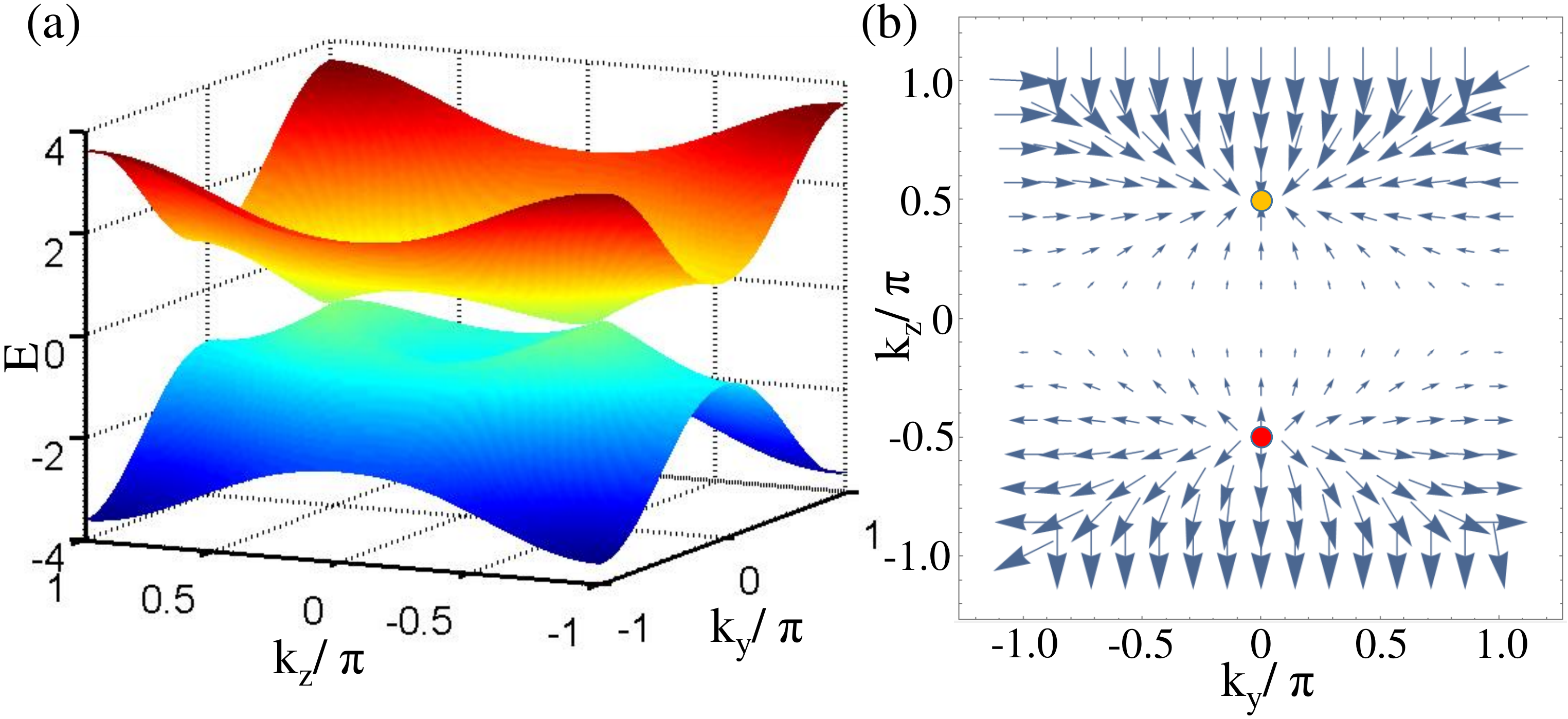}
\caption{(Color online) (a) The band dispersion of the double-Weyl
semimetal in the $k_y$-$k_z$ plane with $k_x=0$ and $m_z=2$. (b)
The vector distribution of the Berry curvature
$\boldsymbol{F}(\boldsymbol{k})$ in the $k_y$-$k_z$ plane. The
double-Weyl point $\boldsymbol{W}_+=(0,0,+\frac{\pi}{2})$ denoted
by yellow dot is a sink in momentum space and the other point
$\boldsymbol{W}_-=(0,0,-\frac{\pi}{2})$ denoted by red dot is a
source in the momentum space.}\label{DoubleWeyl}
\end{figure}

For this lattice system under the periodic boundary condition, the
tight-binding Hamiltonian can be rewritten as
$\hat{H}=\sum_{\boldsymbol{k},\sigma\sigma'}\hat{a}^{\dag}_{\boldsymbol{k}\sigma}[\mathcal{H}(\boldsymbol{k})]_{\sigma\sigma'}\hat{a}_{\boldsymbol{k}\sigma}$,
where
$\hat{a}_{\boldsymbol{k}\sigma}=1/\sqrt{V}\sum_{\boldsymbol{r}}e^{-i\boldsymbol{k\cdot
r}}\hat{a}_{\boldsymbol{r}\sigma}$ is the annihilation operator in
momentum space $\boldsymbol{k}=(k_x,k_y,k_z)$, and
$\mathcal{H}(\boldsymbol{k})=\boldsymbol{d}(\boldsymbol{k})\cdot\boldsymbol{\hat{\sigma}}$
is Bloch Hamiltonian. Here
$\boldsymbol{d}(\boldsymbol{k})=(d_x,d_y,d_z)$ denotes the Bloch
vectors: $d_x = \cos k_x-\cos k_y$, $d_y = \sin k_x\sin k_y$, and
$d_z = m_z-\cos k_x-\cos k_y-\cos k_z$, with the lattice spacing
$a\equiv1$ and $\hbar\equiv1$ hereafter. The Bloch Hamiltonian is
thus given by
\begin{eqnarray} \label{HK}
\nonumber \mathcal{H}(\boldsymbol{k})&=&(\cos k_x-\cos k_y)\sigma_x+\sin k_x\sin k_y\sigma_y\\
&&+(m_z-\cos k_x-\cos k_y-\cos k_z)\sigma_z.
\end{eqnarray}
The energy spectrum of the system is given by $E_{\pm}(\boldsymbol{k})=\pm|\boldsymbol{d}(\boldsymbol{k})|$.
The bulk gap closes when $d_{x}(\boldsymbol{k})=d_{y}(\boldsymbol{k})=d_{z}(\boldsymbol{k})=0$. 
By solving the equations, we can obtain a pair of twofold
degenerate points that are double-Weyl  points
$\boldsymbol{W}_{\pm}=(0,0,\pm\arccos(m_{z}-2))$ for $1<m_z<3$ and
another pair of double-Weyl points
$\boldsymbol{W}_{\pm}=(\pi,\pi,\pm\arccos(m_{z}+2))$ for
$-3<m_z<-1$. For instance, the energy spectrum as a function of
$k_y$ and $k_z$ with fixed $k_x=0$ for $m_z=2$ is shown in  Fig.
\ref{DoubleWeyl}(a), where two double-Weyl points locate at
$(0,0,\pm\pi/2)$.

We consider the nodes
$\boldsymbol{W}_{\pm}=(0,0,\pm\arccos(m_{z}-2))$ to further show
that they are double-Weyl points. Expanding the Bloch Hamiltonian
near the two nodes with
$\boldsymbol{q}=(q_x,q_y,q_z)\equiv\boldsymbol{k}-\boldsymbol{W}_{\pm}$
yields  the low-energy effective Hamiltonian
\begin{equation}
\mathcal{H}_{\pm}\approx\frac{1}{2}(q_y^2-q_x^2)\sigma_x+q_xq_y\sigma_y+\chi v_zq_z\sigma_z,
\end{equation}
where $\chi=\pm1$ respectively for the two nodes
$\boldsymbol{W}_{\pm}$ and $v_z=\sqrt{1-(m_z-2)^{2}}$.  The
effective Hamiltonian shows that the dispersion near the nodes is
quadratic in $k_x$ and $k_y$ and linear in $k_z$. One can rewrite
the low-energy effective Hamiltonian as
\begin{equation}
\mathcal{H}_{\pm}=\epsilon\left(
\begin{array}{cc}
\chi\cos\theta & -\sin\theta e^{i2\varphi}\\
-\sin\theta e^{-i2\varphi} & -\chi\cos\theta\\
\end{array},
\right)
\end{equation}
where $v_{\parallel}=\frac{1}{2}$, $q_{\parallel}^{2}=q_x^{2}+q_y^{2}$,
$\epsilon=\sqrt{(v_zq_z)^{2}+v_{\parallel}^{2}(q_x^{2}+q_y^{2})^{2}}$,
$\cos\theta=v_zq_z/\epsilon$, $\sin\theta=v_{\parallel}q_{\parallel}^{2}/\epsilon$, $\sin\varphi=q_x/q_{\parallel}$,
 and $\cos\varphi=q_y/q_{\parallel}$. The eigenstates of the lowest band near the two nodes with index $\chi=\pm1$ are
 respectively  given by $|u_{0}\rangle=(\sin\frac{\theta}{2}e^{i2\varphi},\cos\frac{\theta}{2})^T$ and $|u_{0}\rangle=(\cos\frac{\theta}{2}e^{i2\varphi},-\sin\frac{\theta}{2})^T$. The Chern number (topological charge) $C_{\chi}$ of the nodes can thus be computed by integrating the Berry curvature over an arbitrary Fermi sphere $S$ that encloses each node \cite{Detect}:
\begin{eqnarray}
C_{\chi}=\frac{1}{2\pi}\oint_{\boldsymbol{S}}d\boldsymbol{S}\cdot\boldsymbol{F}=-2\chi,
\end{eqnarray}
where the Berry curvature
$\boldsymbol{F}=\nabla\times\boldsymbol{A}$ and
$\boldsymbol{A}=(A_{\theta},A_{\phi})$ is  the Berry connection
given by $A_{\theta}=i\langle u_0|\partial_\theta|u_0\rangle=0$
and $A_\varphi=i\langle
u_0|\partial_\varphi|u_0\rangle=-2\chi\sin^{2}\frac{\theta}{2}$.
The above results reveal that the two nodes have opposite
topological charges of $\pm2$ and quadratic dispersion, in
contract to the standard Weyl points in Weyl semimetals that have
topological charges of $\pm1$ and linear dispersion, so they are
named double-Weyl points. Thus the system is in the double-Weyl
semimetal phase when the Fermi level lies in the vicinity of the
double-Weyl points.

In momentum space, the gauge field associated with the Berry
curvature near the neighborhood of Weyl node behaves like a
magnetic field originating from a magnetic monopole. Here the
opposite chirality of the paired double-Weyl points can also be
viewed as a monopole-antimonopole pair in the momentum space. To
show this point, we calculate the Berry curvature as a function of
the momentum $\boldsymbol{k}$:
$\boldsymbol{F}(\boldsymbol{k})=\nabla\times\boldsymbol{A}(\boldsymbol{k})$
with the Berry connection $\boldsymbol{A}(\boldsymbol{k})=i\langle
u_0(\boldsymbol{k})|\nabla_{\boldsymbol{k}}|u_0(\boldsymbol{k})\rangle$
defined by the wave function $|u_0(\boldsymbol{k})\rangle$ in the
lowest band. For the two bands system, the lowest-band Berry
curvature in the momentum space is given by \cite{anomalous}
\begin{eqnarray}
F^{a}=\epsilon_{abc}F_{bc}=\epsilon_{abc}\left[\frac{1}{2d^{3}}\boldsymbol{d}\cdot\left(\frac{\partial\boldsymbol{d}}{\partial k_b}\times\frac{\partial\boldsymbol{d}}{\partial k_c}\right)\right],
\end{eqnarray}
where the three components are obtained as $F^{x}=(\sin k_x\cos
k_x\cos k_y\sin k_z-\sin k_x\sin k_z)/N(\boldsymbol{k})$,
$F^{y}=(\cos k_x\cos k_y\sin k_y\sin k_z-\sin k_y\sin
k_z)/N(\boldsymbol{k})$, and
$F^{z}=(2\sin^{2}k_y+2\sin^{2}k_x\cos^{2}k_y + (\cos
k_z-m_z)(\sin^{2}k_x\cos k_y+\cos
k_x\sin^{2}k_y))/N(\boldsymbol{k})$, with
$N(\boldsymbol{k})=2[(\cos k_x-\cos
k_y)^{2}+\sin^{2}k_x\sin^{2}k_y+(m_z-\cos k_x-\cos k_y-\cos
k_z)^{2}]^{3/2}$. The vector distribution of the Berry curvature
$\boldsymbol{F}(\boldsymbol{k})$ in the $k_x=0$ plane are plotted
in Fig. \ref{DoubleWeyl}(b), which clearly shows that the
double-Weyl points located at
$\boldsymbol{W}_{\pm}=(0,0,\pm\frac{\pi}{2})$ behave as sink and
source of the Berry flux.

Finally in this section, we note that the double-Weyl points are
stabilized by the $C_{4h}$ symmetry of the system, which 
consists of $C_4$ point-group symmetry and mirror symmetry $P$
with the Bloch Hamiltonian obeying \cite{Multi,Shivamoggi}
\begin{eqnarray}
C_4\mathcal{H}(\boldsymbol{k})C_{4}^{-1}=\mathcal{H}(P\boldsymbol{k})
\end{eqnarray}
where $C_4=e^{-i\frac{\pi}{2}\sigma_z}$ is a point-group operator
for the fourfold rotation about the $z$ axis and $P$ is a matrix
transfering $(k_x,k_y,k_z)$ to $(k_y,-k_x,-k_z)$. Here we can
define
$f(\boldsymbol{k})=d_x(\boldsymbol{k})-id_y(\boldsymbol{k})$ and
$\sigma_{\pm}=(\sigma_{x}\pm i\sigma_{y})/2$, such that the Bloch
Hamiltonian of the two-band model can be rewritten as
$\mathcal{H}(\boldsymbol{k})=f(\boldsymbol{k})\sigma_++f^{*}(\boldsymbol{k})\sigma_-+d_z(\boldsymbol{k})\sigma_z$.
Thus the transform of $\mathcal{H}(\boldsymbol{k})$ under $C_4$
leads to
$C_4H(\boldsymbol{k})C_4^{-1}=f(\boldsymbol{k})e^{-i\frac{\pi}{2}\sigma_z}\sigma_+e^{i\frac{\pi}{2}\sigma_z}+f^{*}
(\boldsymbol{k})e^{-i\frac{\pi}{2}\sigma_z}\sigma_-e^{i\frac{\pi}{2}\sigma_z}+d_z(\boldsymbol{k})\sigma_z$.
Since the Bloch vectors of the double-Weyl semimetal satisfy
$f(P\boldsymbol{k})=-f(\boldsymbol{k})$,
$f^{*}(P\boldsymbol{k})=-f^{*}(\boldsymbol{k})$ and
$d_z(P\boldsymbol{k})=d_z(\boldsymbol{k})$, the system preserves
$C_{4h}$ symmetry. When the $C_{4h}$ symmetry are broken, the
double-Weyl points will be destroyed and the system is no longer
in the double-Weyl semimetal phase. Thus it would be valuable to
study the symmetry-breaking effects and the related phase
transition in this tunable system.

\begin{figure}[tbph]
\centering
\includegraphics[width=8.5cm]{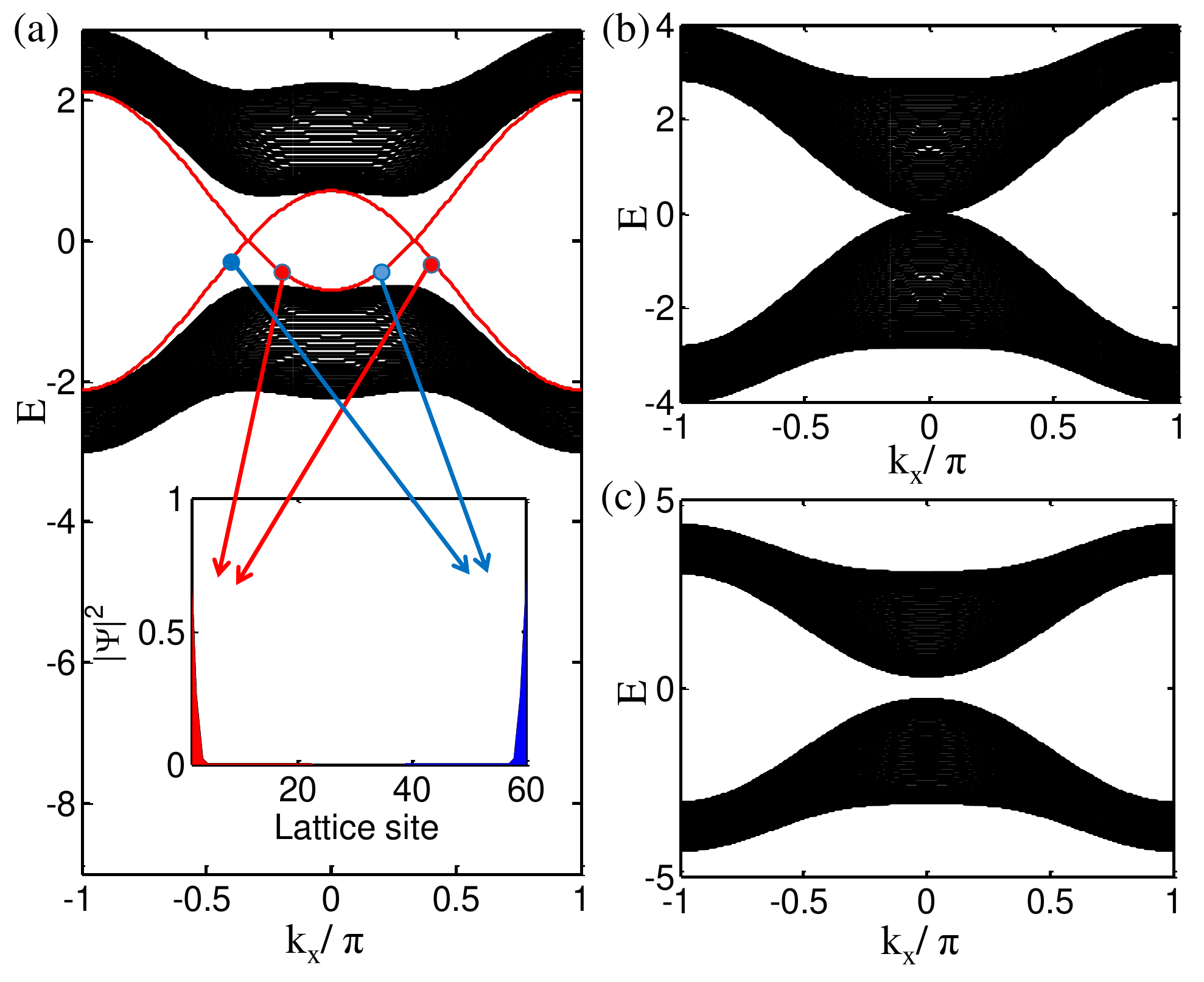}
\caption{(Color online) (a) The energy spectrum and edge states
of the reduced chain with lattice sites $L_y=60$ under open
boundaries for $k_z=0$. There are two chiral states per surface
and the inset shows the density distributions of four typical edge
modes. (b) The energy spectrum for $k_z=0.5\pi$. (c) The energy
spectrum for $k_z=0.6\pi$. Other parameter $m_z=2$.}
\label{ChainEnergy}
\end{figure}

\section{topological properties of the simulated double-Weyl semimetals}

To further study the topological properties of this system, we
consider the Bloch Hamiltonian with dimension reduction method:
considering $k_{z}$ as a good quantum number and then reduce the
three-dimensional system to a set of two-dimensional subsystems
with $k_{z}$ as a parameter. For a fixed $k_{z}$, the reduced
Bloch Hamiltonian $\mathcal{H}_{k_{z}}(k_{x},k_{y})$ is given by
\begin{eqnarray}
\nonumber \mathcal{H}_{k_{z}}(k_{x},k_{y})&=&(\cos k_{x}-\cos k_{y})\sigma_{x}+\sin k_{x}\sin k_{y}\sigma_{y}\\
&&+(M_{z}-\cos k_{x}-\cos k_{y})\sigma_{z},
\end{eqnarray}
where $M_{z}=m_{z}-\cos k_{z}$. The bands of these subsystems with
fixed $k_{z}$ are all gapped  when $k_{z}\neq \pm k_{z}^{c}$ with
$k_{z}^{c}=\arccos(m_{z}-2)$. Under this condition,
$\mathcal{H}_{k_{z}}(k_{x},k_{y})$ describes effective
two-dimensional Chern insulators since the $k_z$-dependent Chern
number is given by
\begin{equation} \label{Ckz}
C_{k_z}=\frac{1}{4\pi} \int_{-\pi}^{\pi}dk_x\int_{-\pi}^{\pi}dk_y~\boldsymbol{\hat{d}}\cdot\left(\partial_{k_x}\boldsymbol{\hat{d}}\times\partial_{k_y}\boldsymbol{\hat{d}}\right),
\end{equation}
where $\boldsymbol{\hat{d}}\equiv
\boldsymbol{d}/|\boldsymbol{d}|$. In the parameter regime
$1<|m_z|<3$, we obtain that $C_{k_z}=2$ for the planes with
$-k_{z}^{c}<k_z<k_{z}^{c}$ and $C_{k_z}=0$ for other cases. With
similar dimension reduction method, such a two-dimensional Chern
insulator can be regarded as a fictitious one-dimensional chain
subjected to an external parameter $k_x$ since we can consider
$k_x$ as a good quantum number. The tight-binding Hamiltonian of
such a one-dimensional chain along the $y$ axis can be written as

\begin{eqnarray}
\hat{H}_{y}(k_x,k_z)&=& \nonumber-\frac{1}{2}\sum_{i_y}\left(\hat{a}^{\dag}_{i_y,\uparrow}\hat{a}_{i_y+1,\uparrow}-\hat{a}^{\dag}_{i_y,\downarrow}\hat{a}_{i_y+1,\downarrow}+\text{H.c.}\right)\\ \nonumber&&-\frac{1}{2}\sum_{i_y}[(1+\sin k_x)\hat{a}^{\dag}_{i_y,\uparrow}\hat{a}_{i_y+1,\downarrow}+\\ \nonumber&&~~~~~~~~(1-\sin k_x)\hat{a}^{\dag}_{i_y,\uparrow}\hat{a}_{i_y-1,\downarrow}+\text{H.c.}]\\
\nonumber&&+\sum_{i_y}M_{zx}(\hat{a}^{\dag}_{i_y,\uparrow}\hat{a}_{i_y,\uparrow}-\hat{a}^{\dag}_{i_y,\downarrow}\hat{a}_{i_y,\downarrow})\\
&&+\sum_{i_y}\cos
k_x(\hat{a}^{\dag}_{i_y,\uparrow}\hat{a}_{i_y,\downarrow}+\hat{a}^{\dag}_{i_y,\downarrow}\hat{a}_{i_y,\uparrow}),
\end{eqnarray}
where $M_{zx}=M_z-\cos k_x=m_z-\cos k_x-\cos k_x$. The
tight-binding Hamiltonian can be used to study the nontrivial
edges states in the system.

We numerically calculate the energy spectrum $E(k_x)$ of the
reduced chain with length $L_y=60$ under open boundary conditions
in different $k_z$ planes for fixed $m_z=2$, which corresponds to
$k_{z}^{c}=0.5\pi$. As shown in Fig. \ref{ChainEnergy}(a) for
$k_z=0$, two symmetric bulk bands with an energy gap accompanies
two chiral in-gap states per surface. The two chiral states have
degeneracies and connect the separated bands, which is consistent
with bulk-edge correspondence in this case with the bulk Chern
number $C_{k_z}=2$. The two chiral states gradually spread into
the bulk when their energies are closer to the bulk bands. The
density distributions of some edge modes are shown in the inset in
Fig. \ref{ChainEnergy}(a) for typical $k_x$. Increasing $|k_z|$ to
the critical points $k_{z}^{c}$, the two degeneracies of surface
states move to the center and then merge at the double-Weyl points
at $k_z=\pm k_{z}^{c}$, with the energy spectrum of
$k_z=k_{z}^{c}=0.5\pi$ being shown in Fig. \ref{ChainEnergy}(b).
When $k_z$ inters the region $|k_z|>k_{z}^{c}$, the energy
spectrum is again gapped but there is no chiral edge state since
$C_{k_z}=0$ in this region, with the case of $k_z=0.6\pi$ shown in
Fig. \ref{ChainEnergy}(c). The change of  topological invariant
$C_{k_z}$ from 2 to 0 indicates a double-Weyl point of monopole
charge 2.

\begin{figure}[tbph]
\centering
\includegraphics[width=8.5cm]{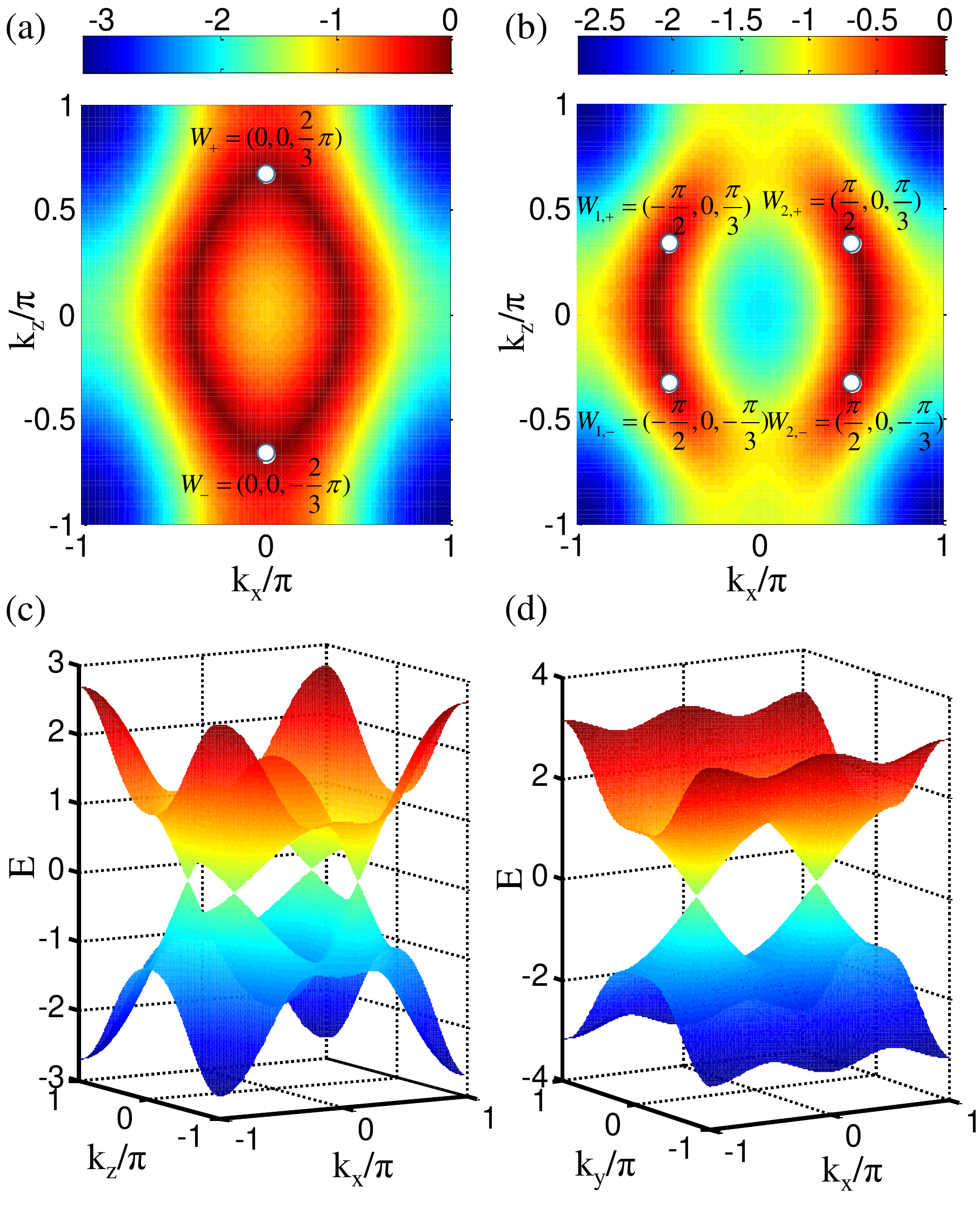}
\caption{(Color online) (a) Fermi arcs connecting the mimicked
double-Weyl points $W_{\pm}$ with monopole charges $\mp2$ denoted
by white dots. The black line denotes Fermi arcs formed by gapless
zero-energy edge modes. (b) Fermi arcs when $\delta=1$. The two
double-Weyl points split into four single-Weyl points when
$\delta\sigma_x$ term is added, and the Fermi arcs (black lines)
terminate at four points denoted by white dots. $W_{1,+}$ and
$W_{2,+}$ have monopole charge $-1$, $W_{1,-}$ and $W_{2,-}$ have
monopole charge $+1$. (c) and (d) The band dispersions $E(k_x,k_z)$
and $E(k_x,k_y)$ for $\delta=1$ with fixed $k_y=0$ and
$k_z=\frac{\pi}{3}$, respectively. The other parameter is $m_z=1.5$.}
\label{FermiArc}
\end{figure}

We further study the Fermi-arc zero modes (with energy $E=0$)  by
numerically calculating the energy spectrum $E(k_x,k_z)$ of the
surface states, with the results being plotted in Fig.
\ref{FermiArc}(a) for typical parameter $m_z=1.5$. In this case,
the two double-Weyl points locate at
$\boldsymbol{W}_{\pm}=(0,0,\pm\frac{2\pi}{3})$, and they are
connecting by two Fermi arcs with $E=0$ plotted with the black
lines. We also find that if $m_z$ are approaching to the critical
values $\pm 1$ or $\pm 3$, the Fermi arcs shrink since the two
double-Weyl nodes move to each other, and they vanish entirely at
the phase boundaries when the two Weyl points merge.

We proceed to study the effects of symmetry breaking in the
double-Weyl semimetals. To this end, we can add a term
$\mathcal{H}_P=\delta \sigma_x$ to the Bloch Hamiltonian in Eq.
(\ref{HK}) to break its $C_4$ symmetry, the resultant Hamiltonian
$\tilde{\mathcal{H}}=\mathcal{H}+\mathcal{H}_P$ becomes
\begin{eqnarray} \label{HKP}
\nonumber \tilde{\mathcal{H}}&=&(\cos k_x-\cos k_y +\delta)\sigma_x+\sin k_x\sin k_y\sigma_y\\
&&+(m_z-\cos k_x-\cos k_y-\cos k_z)\sigma_z.
\end{eqnarray}
In the optical lattice, the $\mathcal{H}_P$ term corresponds  to
the tunable in-site coupling between the two spin states described
by
$H_P=\delta\sum_{\boldsymbol{r}}\hat{a}^{\dag}_{\boldsymbol{r},\uparrow}\hat{a}_{\boldsymbol{r},\downarrow}+\text{H.c.}$,
which can be realized by additional Raman coupling. In this case,
we find that the bulk gap can be closed if $|\delta|\leqslant2$.
In particular, for $1-m_z\leqslant\delta\leqslant3-m_z$, we can
find four single Weyl points located
at$\boldsymbol{W}_{1,\pm}=(-\arccos(1-\delta),0,\pm\arccos(m_z-2+\delta))$
and
$\boldsymbol{W}_{2,\pm}=(\arccos(1-\delta),0,\pm\arccos(m_z-2+\delta))$.
For $m_z+1\leqslant\delta\leqslant m_z+3$, there are also four
single Weyl points at
$(k_x,k_y,k_z)=(\pi,\pm\arccos(1-\delta),\pm\arccos(m_z-2+\delta))$.
Similarly, when $-2\leqslant\delta\leqslant0$, the single Weyl
points are at
$(k_x,k_y,k_z)=(0,\pm\arccos(1+\delta),\pm\arccos(m_z-2+\delta))$
for $m_z-3\leqslant\delta\leqslant m_z-1$ and at
$(k_x,k_y,k_z)=(\pm\arccos(-1-\delta),\pi,\pm\arccos(m_z+2+\delta))$
for $-3-m_z\leqslant\delta\leqslant -1-m_z$.

To reveal the topological nature of these gapless points more
clearly, we first expand the Hamiltonian near the four points
$\boldsymbol{W}_{\lambda,\mu}$ with $\lambda=1,2$ and $\mu=+,-$.
We obtain the corresponding low-energy effective Hamiltonian
$\mathcal{H}_{\lambda,\mu}$ (the other three cases proceed
similarly):
\begin{equation}
\mathcal{H}_{\lambda,\mu}\approx(-1)^{\lambda}\alpha[q_y\sigma_y -(q_x-q_z)\sigma_x]+\mu \alpha_zq_z\sigma_z,
\end{equation}
where $\alpha=\sqrt{1-(1-\delta)^{2}}$  and
$\alpha_z=\sqrt{1-(m_z-2+\delta)^{2}}$ are the effective Fermi
velocities, and
$\boldsymbol{q}=(q_x,q_y,q_z)=\boldsymbol{k}-\boldsymbol{W}_{1,\pm}$
or $\boldsymbol{q}=\boldsymbol{k}-\boldsymbol{W}_{2,\pm}$ for the
four points. Thus, the dispersion near the gapless points is
linear along the three momentum directions, indicating that these
nodes are single-Weyl points. We then consider the evolution of
the Fermi arcs as increasing $\delta$ from 0 to 2, and find that
each double-Weyl point of monopole charge $+2(-2)$ in the
double-Weyl semimetal ($1<m_z<3$ and $\delta=0$) splits into two
pairs of single-Weyl points with monopole charge $+1(-1)$
connected by two disconnected Fermi arcs. When $\delta=2$ the two
pairs of single-Weyl points merge and the Fermi arcs disappear.
For example, there are two Fermi arcs which terminate at the four
single-Weyl points for $\delta=1$ and $m_z=1.5$, as shown in Fig.
\ref{FermiArc}(b). The corresponding energy spectra on the
$k_x-k_z$ ($k_y=0$) and $k_x-k_y$ ($k_z=0$) planes are
respectively shown in Fig.3 (c) and (d), which indicate the linear
dispersion of the four single-Weyl points along each momentum
direction. Therefore in this parameter region ($1<m_z<3$ and
$0<\delta<2$), the system is in the single-Weyl semimetal phase
with four single-Weyl points, which can be obtained for the other
three parameter regions.

\begin{figure}[tbph]
\centering
\includegraphics[width=7cm]{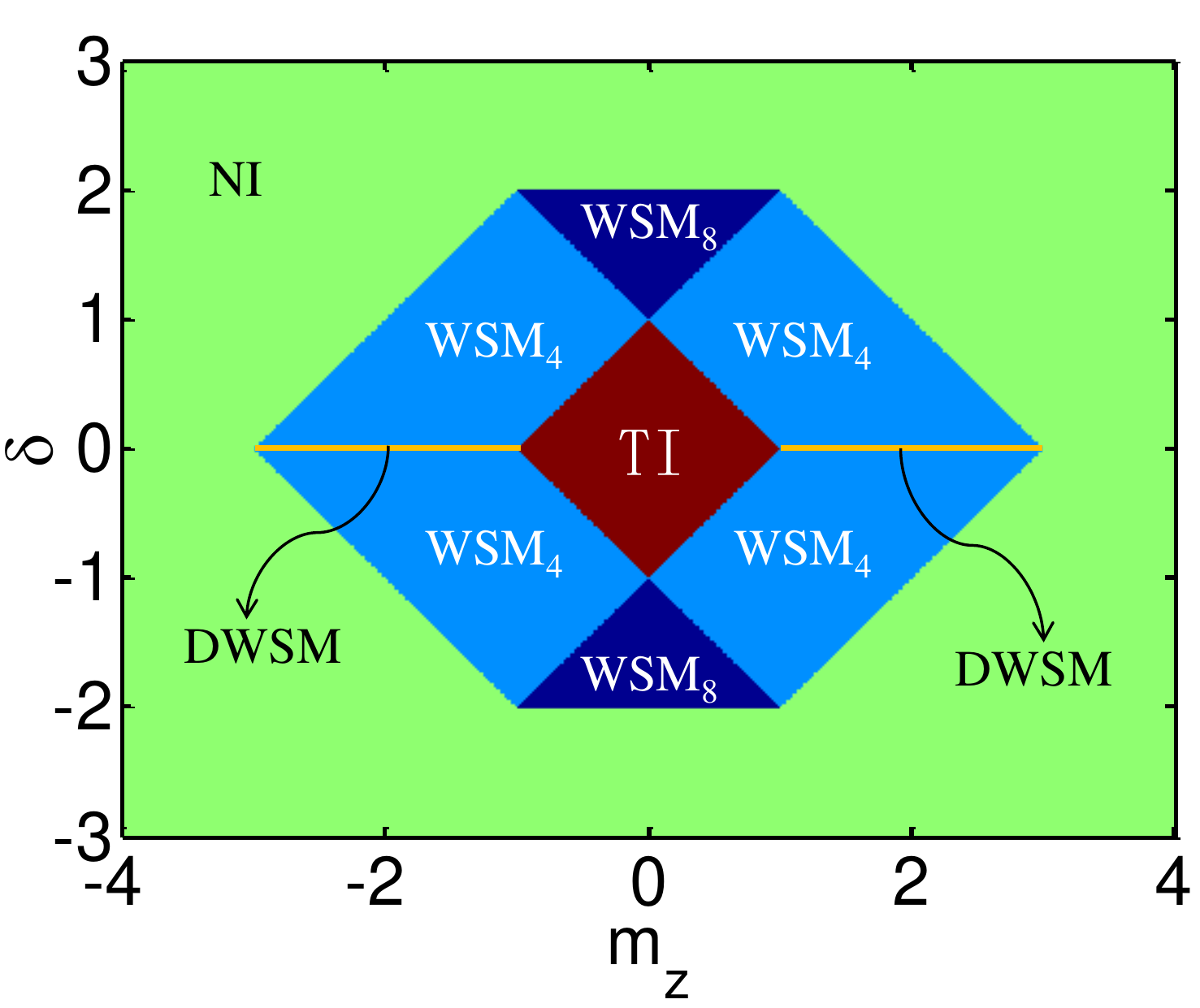}
\caption{(Color online) The phase diagram of the Hamiltonian in
Eq. (\ref{HKP}). TI denotes the topological insulating phase (dark
red), NI denotes the normal band insulating phase (green), WSM$_8$
the Weyl semimetal phase with eight single-Weyl points (dark
blue), WSM$_4$ denotes the Weyl semimetal phase with four
single-Weyl points (blue), and DWSM is the double-Weyl semimetal
phase (yellow lines).} \label{phase-diagram}
\end{figure}

\begin{figure}[tbph]
\centering
\includegraphics[width=8cm]{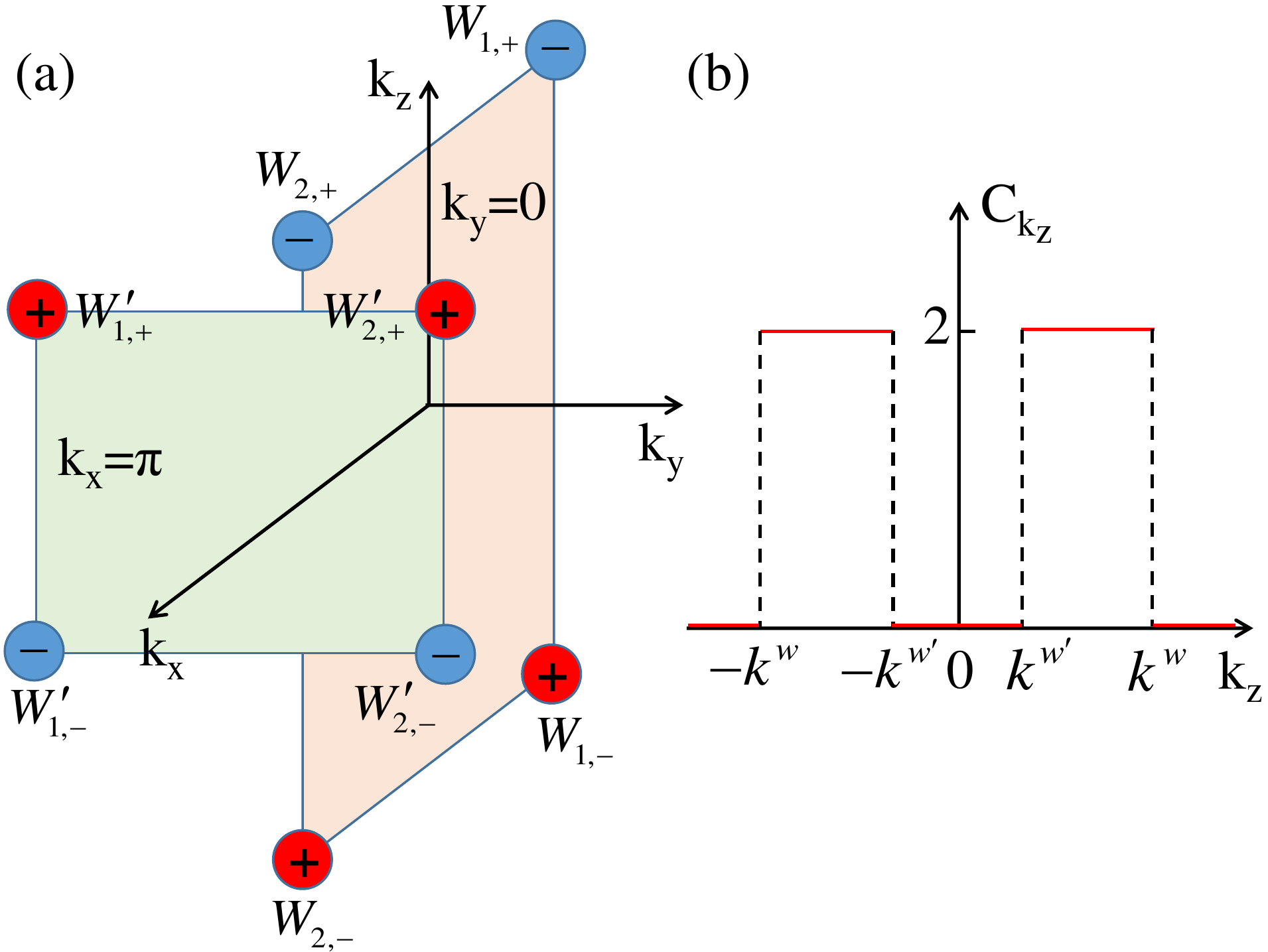}
\caption{(Color online)(a) Illustration of the eight single-Weyl
points. The monopole charges of $\boldsymbol{W'}_{1,-}$,
$\boldsymbol{W'}_{2,-}$, $\boldsymbol{W}_{1,+}$,
$\boldsymbol{W}_{2,+}$ are -1, and the monopole charges of
$\boldsymbol{W'}_{1,+}$, $\boldsymbol{W'}_{2,+}$,
$\boldsymbol{W}_{1,-}$, $\boldsymbol{W}_{2,-}$ are +1. (b) The
Chern number as function of $k_z$. Here
$k^{w}=\arccos(m_z-2+\delta)$ and $k^{w'}=\arccos(m_z+2-\delta)$.}
\label{WSM8}
\end{figure}

By similar analysis of the gapless points and  the topological
properties, we obtain the phase diagram for the Hamiltonian in Eq.
(\ref{HKP}) in the parameter space spanned by $m_z$ and $\delta$,
as shown in Fig. \ref{phase-diagram}. In the phase diagram, apart
from the double-Weyl semimetal phase (denoted by DWSM) and the
single-Weyl semimetal phase with four Weyl points (denoted by
WSM$_4$), there are other three different phases: a normal band
insulating phase (denoted by NI) with $C_{k_z}=0$ when $m_z$ or
$\delta$ is large enough to open a trivial energy gap, a weak
topological insulating phase (denoted by TI) with $C_{k_z}=2$ and
chiral edge states for all the range of $k_z$ when
$-m_z-1<\delta<1-m_z$ and $m_z-1<\delta<m_z+1$, and a single-Weyl
semimetal phase with eight Weyl points (denoted by WSM$_8$). For
the WSM$_8$ phase with $1<\delta<2$, the eight (four pairs)
single-Weyl points locate at
$\boldsymbol{W}_{1,\pm}=(-\arccos(1-\delta),0,\pm\arccos(m_z-2+\delta))$,
$\boldsymbol{W}_{2,\pm}=(\arccos(1-\delta),0,\pm\arccos(m_z-2+\delta))$,
$\boldsymbol{W'}_{1,\pm}=(\pi,-\arccos(\delta-1),\pm\arccos(m_z+2-\delta))$,
and
$\boldsymbol{W'}_{2,\pm}=(\pi,\arccos(\delta-1),\pm\arccos(m_z+2-\delta))$.
Figure \ref{WSM8}(a) depicts the position and monopole charge of
these eight single-Weyl points in momentum space. The
corresponding $k_z$-dependent Chern number $C_{k_z}$ as a function
of $k_z$ is plotted in Fig. \ref{WSM8}(b). We find that
$C_{k_z}=2$ when $k_z\epsilon(-k^{w},-k^{w'})$ with
$k^{w}=\arccos(m_z-2+\delta)$ and $k^{w'}=\arccos(m_z+2-\delta)$.
When $k_z$ sweeps through two single-Weyl points with the total
monopole charge being $+2(-2)$, the Chern number will increase
(decrease) 2, and thus $C_{k_z}=0$ within the region
$(-k^{w'},k^{w'})$. The Chern number increases from 0 to 2 when
$k_z$ sweeps through two points $\boldsymbol{W'}_{1,+}$ and
$\boldsymbol{W'}_{2,+}$, so $C_{k_z}=2$ for
$k_z\epsilon(k^{w'},k^{w})$. Finally $C_{k_z}=0$ when $k_z>k^{w}$
or $k_z<-k^{w'}$.

\section{experimental detection schemes}

At this stage, we have introduced the optical lattice system for
simulation of the double-Weyl semimetal states and explored the
relevant topological properties and the phase diagram. In this
section, we propose practical methods for their experimental
detection. We first show that the simulated Weyl points can
be probed by measuring the atomic Zener tunneling to the excited
band after a Bloch oscillation, and then propose two feasible
schemes to obtain the $k_z$-dependent Chern number from the shift
of hybrid Wannier center and from the the spin polarization in
momentum space, respectively.

\begin{figure}[tbph]
\centering
\includegraphics[width=8.5cm]{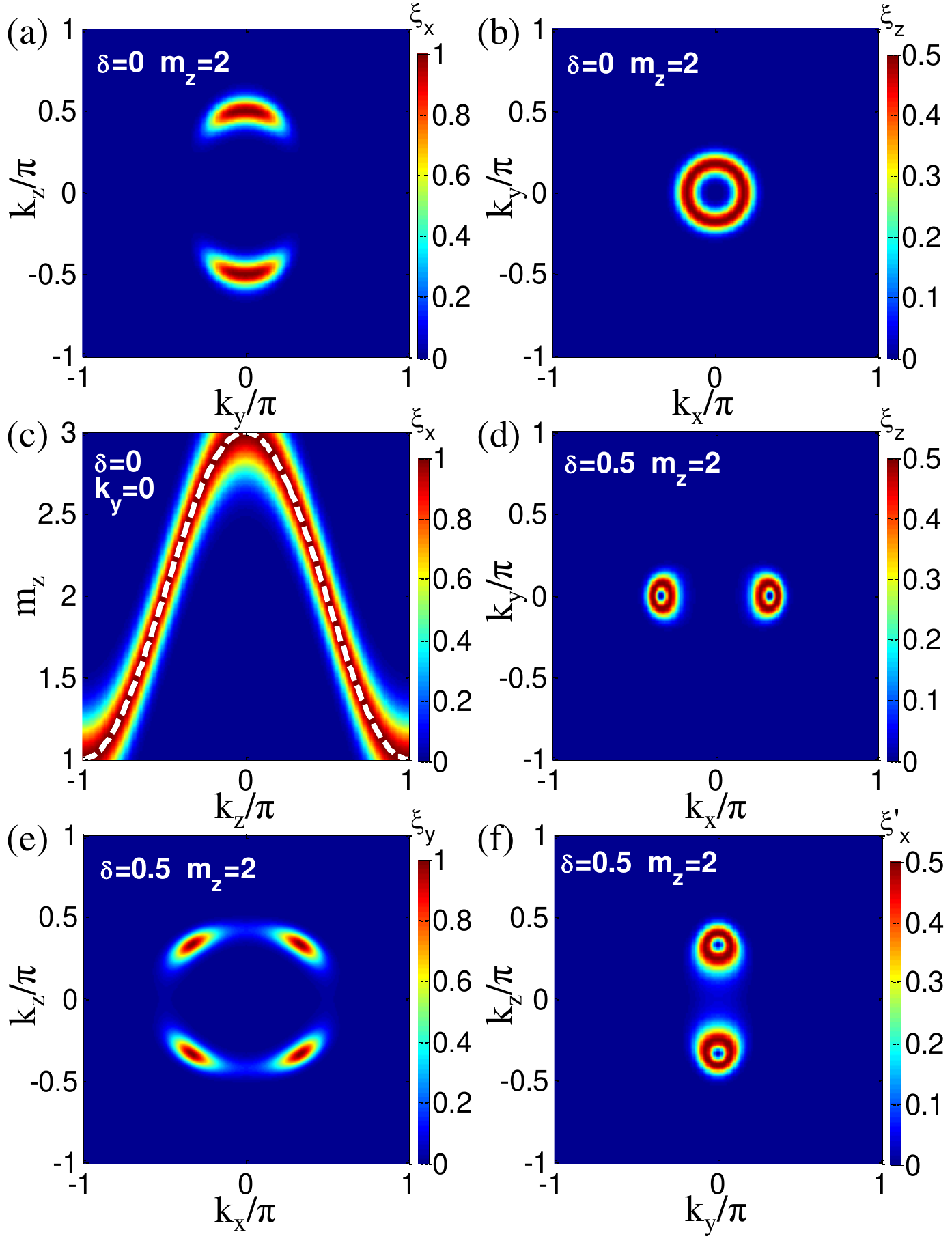}
\caption{(Color online) (a) The distribution $\xi_x(k_y,k_z)$. Two maximum transfer
positions in $k_y$-$k_z$ plane correspond to the positions of the double-Weyl points. (b) The distribution
$\xi_z(k_x,k_y)$. The maximum dip inside the ring profile indicates
$k_x=k_y=0$ for the points. In (a) and (b), $\delta=0$ and $m_z=2$. (c) $\xi_x(k_z)$ for different parameter $m_z$ with $\delta=0$ and fixed $k_y=0$. The maximum transfer positions of $\xi_x(k_z)$ correspond to the expected $k_z$
positions of the paired double-Weyl points, which are denoted by
the white dashed line. (d) The distribution $\xi_z(k_x,k_y)$ with two rings. (e) The distribution $\xi_y(k_x,k_z)$ with four maximum peaks. (f) The distribution $\xi'_x(k_y,k_z)$ with two rings. The patterns in (d-f) with $\delta=0.5$ and $m_z=2$ reveal the positions of four single-Weyl points $(\pm\pi/3,0,\pm\pi/3)$. The other parameter is $F=0.2$ in (a-f).}
\label{LZ}
\end{figure}

\subsection{Detection of the Weyl points}

Here we propose to use the atomic Bloch-Zener oscillation in the
optical lattice to detect the double- and single-Weyl points in this system. One can prepare
noninteracting fermionic atoms in the lower band initially and
apply a constant force $F$ along $\eta$ axis, which push the
atoms moving along $k_{\eta}$ direction in momentum space and
gives rise to Bloch oscillations. Then one can obtain the momentum
distribution of the transfer fraction in the upper band from
time-of-flight measurement after performing a Bloch oscillation.
For the system with the double-Weyl points
$\boldsymbol{W}_{\pm}=(0,0,\pm k_z^c)$, the transfer fractions $\xi_{\eta}$ along different directions
are given by \cite{Tarruell,Lim}
\begin{eqnarray}\label{PLZ}\nonumber
&&\xi_{x}(k_y,k_z)=P_{LZ}^{x}(k_y,k_z),\\
&&\xi_{y}(k_x,k_z)=P_{LZ}^{y}(k_x,k_z),\\\nonumber
&&\xi_{z}(k_x,k_y)=2P_{LZ}^{z}(k_x,k_y)[1-P_{LZ}^{z}(k_x,k_y)],
\end{eqnarray}
where the Landau-Zener transition probabilities are $P_{LZ}^{x}=e^{-\pi\Delta_{x}^{2}(k_y,k_z)/4v_{x}F}$, $P_{LZ}^{y}=e^{-\pi\Delta_{y}^{2}(k_x,k_z)/4v_{y}F}$ and $P_{LZ}^{z}=e^{-\pi\Delta_{z}^{2}(k_x,k_y)/4v_{z}F}$, with $v_{x}=v_{y}=v_{\parallel}$ and the energy gaps $\Delta_x=2E_+(k_x=0,k_y,k_z)$, $\Delta_y=2E_+(k_x,k_y=0,k_z)$ and $\Delta_z=2E_+(k_x,k_y,k_z=k_{z}^{c})$ for the Landau-Zener events along the $k_{\eta}$ directions.

We numerically calculate the transfer fractions $\xi_{\eta}$, with
the results for typical parameter shown in Fig. \ref{LZ}. For the
case of $\delta=0$ and $m_z=2$ in Fig. \ref{LZ}(a), there are two maximum
transfer positions of quasimomentum distribution $\xi_x(k_y,k_z)$
in the $k_y$-$k_z$ plane. The positions correspond to the expected
$k_y=0$ and $k_z=\pm k_z^c$ for the paired double-Weyl points. As the energy gap decreases, the transition probability in a Landau-Zener tunnelling increases exponentially. When the band gap closes, the transfer fraction $\xi_x(k_y,k_z)$
will increase dramatically and thus the band crossing points can be clearly identified in
experiments. Due to the $C_4$ symmetry of the Bloch Hamiltonian when $\delta=0$, one can find that $\Delta_y(k_x,k_z)=\Delta_x(k_y,k_z)$ and the distribution of the transfer fraction $\xi_y(k_x,k_z)$ is the same as the one of $\xi_x(k_y,k_z)$ shown in Fig. \ref{LZ}(a) by replacing $k_y$ with $k_x$. When the atoms move along $k_z$ direction, there are
two subsequent Landau-Zener tunnelling. In Fig. \ref{LZ}(b), the
distribution of the transfer fraction $\xi_z(k_x,k_y)$ shows the
ring-type profile. The position with the value $\xi_z=0$ indicates
$k_x=k_y=0$ for the double-Weyl points. In Fig. \ref{LZ}(c), for $\delta=0$ and
fixed $k_y=0$, the maximum transfer positions of quasimomentum
distributions $\xi_x(k_z)$ correspond well to the expected $k_z$
positions of the mimicked double-Weyl points as plotted by the
dashed line. Therefore, by probing the momentum distribution
$\xi_x(k_z,k_y)$ and $\xi_z(k_x,k_y)$ from the standard
time-of-flight measurement after a Bloch oscillation, the
positions of the double-Weyl points in momentum space can be well
revealed.

The method is applicable for detecting the single-Weyl points created by the symmetry breaking when $\delta\neq0$ in this system. We consider the typical case of $\delta=0.5$ and $m_z=2$, with four single-Weyl points in $k_x$-$k_z$ plane. In this case, there are two subsequent Landau-Zener transitions along $k_x$ direction and thus the transfer fraction $\xi_{x}$ in Eq. (\ref{PLZ}) becomes
\begin{eqnarray}
\xi^{'}_{x}(k_y,k_z)=2P^{x}_{LZ}(k_y,k_z)[1-P^{x}_{LZ}(k_y,k_z)],
\end{eqnarray}
while $\xi_{y}$ and $\xi_{z}$ remain the same expressions. The numerical results of $\xi_{z}$, $\xi_{y}$ and $\xi'_{x}$ are respectively shown in Figs. \ref{LZ}(d,e,f). One can find that both the distributions $\xi_z(k_x,k_y)$ and $\xi'_x(k_y,k_z)$ have two rings and the positions inside each ring with $\xi_z=\xi'_x=0$ indicate four band crossing points located at $(\pm\frac{\pi}{3},0,\pm\frac{\pi}{3})$ as expected for the single-Weyl points in this case. The four peaks of transfer fraction $\xi_y(k_x,k_z)$ shown in Fig. \ref{LZ}(e) also reveal the exact positions of the four gapless points. We note that the Bloch-Zener method can not tell the trivial (accidental) gapless points and the non-trivial Weyl points in the bands. However, the double- and single-Weyl points in our model system can be distinguished from the different patterns of $\xi_{\eta}$, as shown in Fig. \ref{LZ}. To detect the topology of the gapless points, one may further perform the interference between two atomic gases traveling across the points in momentum space revealed by the Bloch-Zener method to extract the Berry phases and thus the Chern numbers \cite{Duca}. Below we present two different approaches to measure the band topology in our model system.

\begin{figure*}[tbph]
\centering
\includegraphics[width=14cm]{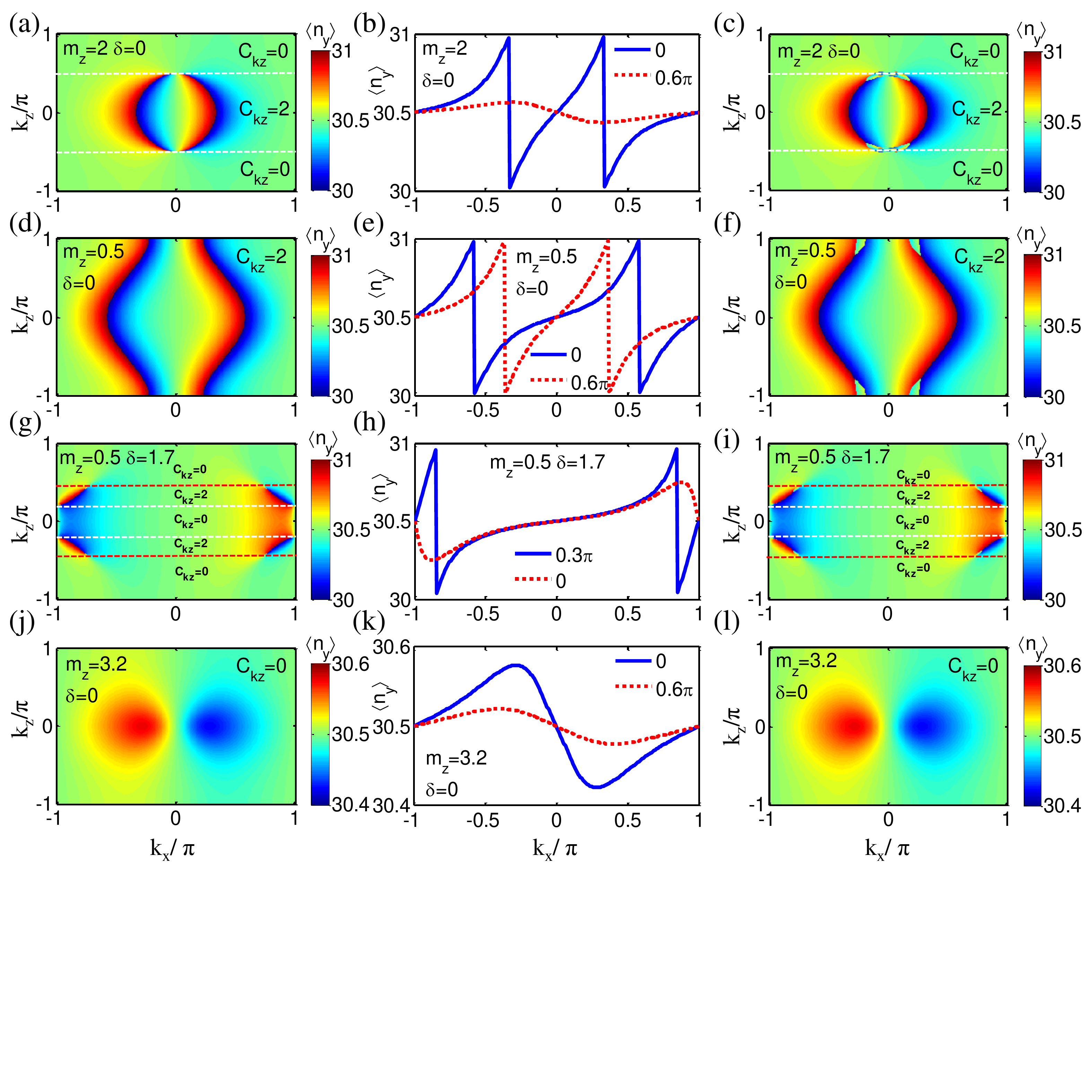}
\caption{(Color online) The hybrid Wannier center in a
tight-binding chain of length $L_y=60$ under the open boundary condition at half filling as a
function of the adiabatic pumping parameter $k_x$ for varying
$k_z$. (a) The profile of the hybrid Wannier center $\langle
n_y(k_x)\rangle$ without trapping potential shows two jumps of
one-unit-cell for $k_z$ (outside) within the region
$(-0.5\pi,0.5\pi)$, with two typical examples $k_z=0$ and
$k_z=0.6\pi$ shown in (b). (c) The profile $\langle
n_y(k_x)\rangle$ under a weak harmonic trap with $V_t=3\times10^{-4}$. The parameters are $\delta=0$ and $m_z=2$ in (a-c).
(d) The profile without trapping potential shows two jumps of $\langle n_y(k_x)\rangle$ for all
$k_z$, with two typical examples shown in (e). (f) The profile under
a harmonic trap with $V_t=6\times10^{-4}$. The parameters are $\delta=0$ and $m_z=0.5$ in (d-f).
(g) The profile without trapping potential exhibits two discontinuous jumps of one-unit-cell when
$k_z\epsilon(-0.46\pi,-0.20\pi)$ or $k_z\epsilon(0.20\pi,0.46\pi)$, with two typical examples shown in
(h). (i) The profile under a harmonic trap with $V_t=2\times10^{-4}$. The parameters are $\delta=1.7$ and $m_z=0.5$ in (g-i).
(j) The profile without trapping potential shows no jump
of $\langle n_y(k_x)\rangle$ for all $k_z$, with two typical
examples shown in (k). (l) The profile under a harmonic
trap with $V_t=3\times10^{-4}$. The parameters are $m_z=3.2$ and $\delta=0$ in (j-l). The $k_z$-dependent Chern numbers
$C_{k_{z}}$ in (a-l) are also plotted.} \label{HWF}
\end{figure*}

\subsection{Detection of the Chern number from the shift of hybrid Wannier center}

We now proceed to propose a realistic scheme to directly measure
the Chern number of the double-Weyl semimetals and other topological states in
optical lattices, based on the particle pumping approach and
hybrid Wannier functions in the band theory
\cite{Pumping1,Pumping2,Pumping3,Wanglei,Marzari,Vanderbilt}. With the dimension reduction
method, the three-dimensional system can be treated as a
collection of $k_z$-modified two-dimensional Chern insulators with the $k_z$-dependent Chern number defined in $k_x$-$k_y$
plane as different slices of out-of-plane quasimomentum $k_z$.
Such a two-dimensional insulating subsystem can be further viewed
as a fictitious one-dimensional insulator subjected to an external
parameter $k_x$. Thus, its Chern number can be defined by the
polarization $P(k_x,k_z)=\frac{1}{2\pi}\int_{-\pi}^{\pi}
\boldsymbol{A}(\boldsymbol{k})dk_y$ for the geometry of the
underlying band structure. According to the modern theory of
polarization \cite{Marzari,Vanderbilt}, the Chern number defined
in $k_x$-$k_y$ space can be obtained from the change in
polarization induced by adiabatically changing the parameter $k_x$
by $2\pi$: $C_{k_z}=\int_{-\pi}^{\pi}\frac{\partial
P(k_x,k_z)}{\partial k_x}dk_x$. For measuring $P(k_x,k_z)$, one
can use another fact that the polarization can alternatively
written as the center of mass of the Wannier function constructed
for the single occupied band.

In this system, the polarization $P(k_x,k_z)$ can be expressed by
means of the centers of the hybrid Wannier functions, which are
localized in the $y$ axis retaining Bloch character in the $k_x$
and $k_z$ dimensions. The variation of the polarization and thus
the Chern number are directly related to the shift of the hybrid
Wannier center along the $y$ axis in the lattice. The shift of
hybrid Wannier center by adiabatically changing $k_x$ is
proportional to the Chern number, which is a manifestation of
topological pumping with $k_x$ being the adiabatic pumping
parameter. In this system, the hybrid Wannier center of a
one-dimensional insulating chain along $y$ axis described by the
Hamiltonian $\tilde{H}=H+H_P$ can be written as
\begin{eqnarray}
\langle n_y(k_x,k_z)\rangle=\frac{\sum_{i_y}i_y\rho_{i_y}(k_x,k_z)}{\sum_{i_y}\rho_{i_y}(k_x,k_z)},
\end{eqnarray}
where $\rho_{i_y}(k_{x},k_{z})$ denotes the density distribution of the hybrid Wannier function as a function of the parameter $k_{x}$ and $k_{z}$ with $i_{y}$ being the lattice-site index in the one-dimensional chain, and takes the following form
\begin{eqnarray}
\rho_{i_y}(k_x,k_z)=\sum_{\text{occ}}|k_x,k_z\rangle_{i_y}{}_{i_y}\langle k_x,k_z|,
\end{eqnarray}
where $|k_{x},k_{z}\rangle_{i_y}$ denotes the hybrid wave function of the system at site $i_y$ and the notation occ denotes the occupied states. In cold atom experiments, the atomic density $\rho_{i_y}(k_{x},k_{z})$ can be directly measured by the hybrid time-of-flight images, which is referring to an \textit{in situ} measurement of the density distribution of the atomic cloud in the $y$ direction during free expansion along the $x$ and $z$ directions. In the measurement, the optical lattice is switched off along the $x$ and $z$ directions while keeping the system unchanged in the $y$ direction. One can map out the crystal momentum distribution along $k_x$ and $k_z$ in the time-of-flight images and a real space density resolution in the $y$ direction can be done at the same time. Thus one can directly extract the Chern number from this hybrid time-of-fight images in the cold atom system.

To demonstrate the feasibility of the proposed method, we numerically calculate $\langle n_y(k_x,k_z)\rangle$ in a tight-binding chain of length $L_y=60$ under the open boundary condition at half filling for some typical parameters, with the results shown in Fig. \ref{HWF}. For $\delta=0$ and $m_z=2$ in Figs. \ref{HWF}(a) and \ref{HWF}(b), the system is in the double-Weyl semimetal phase with $k_{z}^{c}=\pm\pi/2$, and we find that as $k_x$ changing from $-\pi$ to $\pi$, $\langle n_y(k_x)\rangle$ exhibits two discontinuous jumps of one unit cell within the region $k_z\epsilon(-0.5\pi,0.5\pi)$ and the jumps disappear outside this region. To be more clearly, we also plot $\langle n_y(k_x)\rangle$ for $k_z=0$ and $k_z=0.6\pi$ as two examples in Fig. \ref{HWF}(b). The double one-unit-cell jumps driven by $k_x$ indicates that two particles is pumped across the system, as expected for $\mathcal{C}_{k_z}=2$. For $\delta=0$ and $m_z=0.5$ in Figs. \ref{HWF}(d) and \ref{HWF}(e), we find that two discontinuous jumps of $\langle n_y(k_x)\rangle$ for all $k_z$ as $k_x$ changing from $-\pi$ to $\pi$, indicating that the system is in the topological insulating phase with $C_{k_z}=2$. For $m_z=0.5$ and $\delta=1.7$ in Figs. \ref{HWF}(g) and \ref{HWF}(h), the system is in the Weyl semimetal phase with eight single-Weyl points and we find that when $k_z\epsilon(-0.46\pi,-0.20\pi)$ and $k_z\epsilon(0.20\pi,0.46\pi)$, $\langle n_y(k_x)\rangle$ shows two discontinuous jumps of one-unit-cell by varying $k_x$ from $-\pi$ to $\pi$, consistent with $C_{k_z}=2$ in these $k_z$ regions as shown in Fig. \ref{WSM8}. When the system is in the normal band insulating phase for $m_z=3.2$ and $\delta=0$, as expected, there is no jump of the hybrid Wannier center $\langle n_y(k_x)\rangle$ for all $k_z$ by changing adiabatic pumping parameter $k_x$ from $-\pi$ to $\pi$, as shown in Figs. \ref{HWF}(j) and \ref{HWF}(k). We also obtain the results of $\langle n_y(k_x)\rangle$ when the system is in the Weyl semimetal phase with four single-Weyl points, similar with those in Fig. \ref{HWF}(a). This establishes a direct and clear connection between the shift of the hybrid Wannier center and the topological invariant of the system in different phases.

In order to simulate the realistic experiment, we add a weak harmonic trap to this finite-site lattice with the open boundary. The trapping potential in the chain can be effectively described as $H_t=V_t\sum_{i_y}(i_y-\frac{L_y}{2})^{2}\hat{a}^{\dag}_{i_y}\hat{a}_{i_y}$, where $V_t$ is the trap strength. Within a local density approximation, as long as the lower band is still filled at the center of the trap, the shifts of the hybrid Wannier center can be expected to be nearly the same as those without the trap potential. If the band gap $E_g<V_t(i_y-\frac{L_y}{2})^{2}$, the lower band is only partially filled near the two edges and then this pumping argument is no longer well applicable. In practical experiments, one can turn the trap strength to $V_t\sim4E_g/L_y^{2}$ or emphasize the shift of hybrid Wannier center in the central region. With numerical simulations, we demonstrate that the results of $\langle n_y(k_x,k_z)\rangle$ preserve with a deviation less than $2\%$ except the regions near the band crossing points for $V_t=3\times10^{-4}$ in Fig. \ref{HWF}(c) and \ref{HWF}(l), $V_t=6\times10^{-4}$ in Fig.  \ref{HWF}(f), and $V_t=2\times10^{-4}$ in Fig. \ref{HWF}(i). They are consistent with the estimates in the local-density analysis.

\subsection{Detection of the band topology from the spin polarization in momentum space}

Below we propose an alternative method to probe the
band topology of the $C_4$-symmetric Bloch Hamiltonian with $\delta=0$ from the
spin polarization in momentum space, which can be implemented with
bosonic atoms in the optical lattice. When the system has $C_4$ symmetry in the $xy$ plane, we can treat $k_z$ as an
effective parameter and reduce it to a collection of
effective two-dimensional subsystems, whose Chern number $C_{k_z}$
for a fixed $k_z$ can be determined by the following equation \cite{Multi,Shivamoggi}

\begin{eqnarray} \label{symmetry}
e^{i\frac{\pi}{2}C_{k_z}}=\prod_{n\epsilon \text{occ}}\gamma_n(0,0,k_z)\gamma_n(\pi,\pi,k_z)\chi_n(0,\pi,k_z).
\end{eqnarray}
Here $\gamma_n$ and $\chi_n$ are the $C_4$ and $C_2$ eigenvalues on the $n$-th Bloch band at high-symmetry momentum points in $k_x$-$k_y$ plane, respectively.

For our two bands system, $C_4=e^{-i\frac{\pi}{2}\sigma_z}=-i\sigma_z$ and
$C_2=C_{4}^{2}=-1$, such that the term $\chi_n(0,\pi,k_z)=-1$ in Eq.
(\ref{symmetry}) can be dropped from the expression.
The Chern number $C_{k_z}$ of the lower band for different $k_z$ can thus be
determined by the simple relation
\begin{eqnarray} \label{symmetry-Ckz}
e^{i\frac{\pi}{2}C_{k_z}}=S_-(0,0,k_z)S_-(\pi,\pi,k_z)
\end{eqnarray}
where $S_-(0,0,k_z)$ and $S_-(\pi,\pi,k_z)$ are the eigenvalues of
the $\sigma_z$ operator on the lower band. Thus to obtain
$C_{k_z}$ for a given $k_z$, one only needs to measure the
eigenvalues of $\sigma_z$ in the two high  symmetry points in
$k_x$-$k_y$ plane
${\boldsymbol{\Lambda_i}}=\{\boldsymbol{\Gamma}=(0,0,k_z),\boldsymbol{M}=(\pi,\pi,k_z)\}$.
This can simplify the experimental detection of the topological
invariant of the Bloch bands. The high symmetry points
$\boldsymbol{\Lambda_i}$ satisfy that
$P\boldsymbol{\Lambda_i}=\boldsymbol{\Lambda_i}$. Thus, the
constraints at these points give
$f(\boldsymbol{\Lambda_i})=-f(\boldsymbol{\Lambda_i})$ and
$f^{*}(\boldsymbol{\Lambda_i})=-f^{*}(\boldsymbol{\Lambda_i})$,
which imply that $f(\boldsymbol{k})$ and $f^{*}(\boldsymbol{k})$
vanish. So at the high symmetry points $\boldsymbol{\Lambda_i}$,
the Bloch Hamiltonian can be written as
\begin{eqnarray}
\mathcal{H}(\boldsymbol{\Lambda_i})=d_z(\boldsymbol{\Lambda_i})\sigma_z,
\end{eqnarray}
where the energy of the two bands
$E_{\pm}(\boldsymbol{\Lambda_i})=\pm|d_z(\boldsymbol{\Lambda_i})|$.
Since  the Bloch Hamiltonian commutes with the symmetry operator,
i.e., $[C_4,\mathcal{H}(\boldsymbol{\Lambda_i})]=0$, the Bloch
states of the two bands $|u_{\pm}(\boldsymbol{\Lambda_i})\rangle$
are also the eigenstates of $C_4$. Therefore, one can obtain the
Chern number of the lower band for different $k_z$ by measuring
the spin polarization $\langle\sigma_z\rangle$ near the high
symmetry points in momentum space, which can be written as
\begin{equation}\label{polarization}
\langle\sigma_z(\boldsymbol{\Lambda_i})\rangle=\frac{n_\uparrow(\boldsymbol{\Lambda_i})-n_\downarrow(\boldsymbol{\Lambda_i})}{n_\uparrow(\boldsymbol{\Lambda_i})+n_\downarrow(\boldsymbol{\Lambda_i})}.
\end{equation}
Here $n_{\uparrow,\downarrow}(\boldsymbol{\Lambda_i})$ denotes the atomic density of spin states $|\uparrow,\downarrow\rangle$ at the high symmetry points in $k_x$-$k_y$ plane for a fixed $k_z$. Since this detection protocol only requires measurement of the atomic density distribution in momentum space, it can be applied to bosonic atoms, typically Bose-Einstein condesates, in the optical lattice system with the topological bands.

In the experiment with a condensate in the optical lattice, the spin polarization at the two high symmetry momenta can be written as
\begin{equation}
\langle\sigma_z(\boldsymbol{\Lambda_i})\rangle\approx S_-(\boldsymbol{\Lambda_i})f(E_-,T)
+S_+(\boldsymbol{\Lambda_i})f(E_+,T),
\end{equation}
where $f(E_{\pm},T)=1/[e^{(E_{\pm}(\boldsymbol{\Lambda_i})-\mu)/k_BT}-1]$ is the Bose-Einstein statistics with $\mu$ and $T$ respectively being the chemical potential and temperature, and $S_{\pm}(\boldsymbol{\Lambda_i})$ are the eigenvalues of $\sigma_z$ on the lower and upper bands at $\boldsymbol{\Lambda_i}$. Since $S_+(\boldsymbol{\Lambda_i})=-S_-(\boldsymbol{\Lambda_i})$, one has
$\langle\sigma_z(\boldsymbol{\Lambda_i})\rangle\approx S_-(\boldsymbol{\Lambda_i})[f(E_-,T)-f(E_+,T)]$. Thus by preparing a cloud of bosonic atoms with the temperature satisfying $f(E_-(\boldsymbol{\Lambda_i}),T)>f(E_+(\boldsymbol{\Lambda_i}),T)$, one can obtain
\begin{eqnarray}
\text{sgn}[\langle\sigma_z(\boldsymbol{\Lambda_i})\rangle]=\text{sgn}[S_-(\boldsymbol{\Lambda_i})].
\end{eqnarray}
Therefore, the spin polarization $\langle\sigma_z(\boldsymbol{\Lambda_i})\rangle$ can be precisely measured with a condensate at low temperature.

In practical experiments, one can prepare the atoms in the spin-up state and adiabatically load the condensate into the $\boldsymbol{\Lambda_i}$ points. Then one can perform the spin-resolved time-of-flight expansion, which projects the Bloch states onto free momentum states according to the plane-wave expansion with a complete basis of plane waves $\{\psi_{m,n}^{\uparrow}(\boldsymbol{\Lambda_i}),\psi_{p,l}^{\downarrow}(\boldsymbol{\Lambda_i})\}$. The Bloch state of lowest band can be expressed as
$|u_-(\boldsymbol{\Lambda_i})\rangle=\sum_{m,n}a_{m,n}\psi_{m,n}^{\uparrow}(\boldsymbol{\Lambda_i})|\uparrow\rangle+\sum_{p,l}b_{p,l}\psi_{p,l}^{\downarrow}(\boldsymbol{\Lambda_i})|\downarrow\rangle$,
where $a_{m,n}$ and $b_{p,l}$ are coefficients. The spin polarization for the Bloch eigenstates of the lower band at high symmetry points is given by
$\langle\sigma_z(\boldsymbol{\Lambda_i})\rangle=\langle u_-(\boldsymbol{\Lambda_i})|\sigma_z|u_-(\boldsymbol{\Lambda_i})\rangle
 =\sum_{m,n}|a_{m,n}\psi_{m,n}^{\uparrow}(\boldsymbol{\Lambda_i})|^{2}-\sum_{p,l}|b_{p,l}\psi_{p,l}^{\downarrow}(\boldsymbol{\Lambda_i})|^{2}$,
 which gives rise to the expression in Eq. (\ref{polarization}). Finally one can obtain $n_{\uparrow,\downarrow}(\boldsymbol{\Lambda_i})$
 by the time-of-flight measurement, and thus obtain the $k_z$-dependent Chern number of the Bloch bands from Eq. (\ref{symmetry-Ckz}) with $S_-(\boldsymbol{\Gamma})S_-(\boldsymbol{M})=\text{sgn}[\langle\sigma_z(\boldsymbol{\Gamma})\rangle]\text{sgn}[\langle\sigma_z(\boldsymbol{M})\rangle]$ in this case.

Finally, the fact that the topology of the
$C_4$-symmetric bands can be determined by only the Bloch states
at the symmetric momenta can greatly simplify the experimental
detection of the topological bands. Similar protocol has been
implemented to detect the topology of the inversion-symmetric
bands with Bose-Einstein condensates in two-dimensional optical
lattices \cite{2DSOC}. In our three-dimensional system, one can
extract $C_{k_z}$ from the proposed measurements for various $k_z$
and $m_z$, corresponding to the line of $\delta=0$ in the phase
diagram. If all two-dimensional  slices have $C_{k_z}=0$, the
system is in the trivial insulator phase, while it is in the
topological insulator phase if $C_{k_z}=2$ for all $k_z$. The
change of $C_{k_z}$ by two along $k_z$ axis indicates the presence
of double-Weyl points and the system is in the double-Weyl
semimetal phase.

\section{Conclusions}

In summary, we have proposed an optical lattice system for
simulation and exploration of double-Weyl semimetals. We have
investigated the topological properties of the double-Weyl
semimetal phase and obtained a rich phase diagram with several other
quantum phases, which include a topological insulator phase and
two single-Weyl semimetal phases. Furthermore, with numerical
simulations, we have proposed practical methods for the
experimental detection of the mimicked Weyl points and the
characteristic topological invariants with cold atoms in the lattice system.
The proposed system would provide a promising platform for
elaborating the intrinsic exotic physics of double-Weyl semimetals
and the related topological phase transitions that are elusive in
nature.

\acknowledgements{This work was supported by the NKRDP of China
(Grant No. 2016YFA0301803), the NSFC (Grants No.
11604103, No. 11474153, and No. 91636218), the NSF of Guangdong Province (Grant
No. 2016A030313436), and the Startup Foundation of SCNU.}

\end{document}